\documentclass{appolb}
\usepackage{graphicx}
\usepackage{amsmath}
\usepackage{colortbl} % for rowcolor etc.
\usepackage{color}
\usepackage{hyperref}
\hypersetup{colorlinks=false, linkcolor=black, filecolor=black, urlcolor=black}

\allowdisplaybreaks
\newcommand{\bea}{\begin{eqnarray}} % \newcommand{\bea} {\begin{eqnarray*}}
\newcommand{\eea}{ \end{eqnarray}} % \newcommand{\end{eqnarray}} {\end{eqnarray*}  }
\newcommand{\ba}{\begin{eqnarray*}}
\newcommand{\ea}{  \end{eqnarray*}}

\newcommand{\beq}{\begin{equation*} }
\newcommand{\eeq}{\end{equation*}  }

\newcommand{\bqa}{\begin{eqnarray*} }
\newcommand{\eqa}{\end{eqnarray*}}

\begin{document}  \sloppy
\title{S-matrix approach to the Z resonance\footnote{The summary of the material presented at the conference 
``Matter to the Deepest'' in September 2015 at Ustron; also at the FCC-ee meeting in February 2015 at 
Pisa~\cite{RiemannT-Fccee:2015}, at the CALC conference in July 2015 at JINR, Dubna~\cite{RiemannT-calc:2015}, 
and at the HEPKIT workshop in October 2015 at the KIT, Karlsruhe~\cite{RiemannT-hepkit:2015}.} 
%https://indico.cern.ch/event/369827/timetable/#all.detailed}
}
\author{%
{Tord Riemann%
\thanks{E-mail: tordriemann@gmail.com}}
\address{K\"onigs Wusterhausen, Germany}
%\address{Deutsches Elektronen-Synchrotron, DESY,\\ Platanenallee 6, 15738 Zeuthen, Germany}
}

\maketitle
\begin{abstract}
The proposed  $e^+e^-$-collider FCC-ee aims at an unprecedented accuracy for 
$e^+e^-$ collisions into fermion pairs at the $Z$ peak, based on about $10^{13}$ events. 
The S-matrix approach to the $Z$ boson line shape allows the model-independent quantitative description of the reaction $e^+e^- \to {\bar 
f}f$ around the $Z$ peak in terms of few parameters, among them the  mass $M_Z$ and width $\Gamma_Z$ of the $Z$-boson. While weak and 
strong 
corrections remain ``black'', a careful theoretical description of the photonic 
interactions is mandatory. 
I introduce the method and describe applications and the analysis tool SMATASY/ZFITTER.
\end{abstract}
\PACS{11.80.Cr, 12.38.Bx}
  
\section{Introduction}
%\cite{Freitas-Fccee:2015} \cite{Freitas:2014hra}
% 
% 
% Present interest in precision approaches to the $Z$ boson
% 
% 
%  $\sqrt{s}\sim 10$ GeV at  {\bf Belle-II} will measure $10^9$ $\mu^+\mu^-$ events
% 
% See e.g. T.Ferber~\cite{Ferber:dfg2015}.
% 
% See also my talk on topfit, Paris
The FCC-ee project aims at about $10^{13}$ events in the reaction
\bea\label{eq1}
e^+e^- \to (\gamma, Z) \to f^+f^- ~+~ (n~ \gamma)
\eea
at the  $Z$-boson  resonance peak.
% $\to$ Need {\bf complete electroweak 2-loop calculation}; s
% See e.g. A. Freitas~\cite{Freitas-Fccee:2015}.
% 
% Much work on weak two-loop contributions to the $Z$ resonance has been done by Hollik et al., Czakon et al., Freitas et al.
% 
% 
% see~\cite{Freitas:2014hra} and many refs. therein.
The analysis in the Standard Model (SM) will deserve 2-loop accuracy; see~\cite{Freitas:2014hra} and many references therein.
A promising model-independent alternative is the S-matrix 
approach~\cite{Leike:1991pq,Riemann:1992gv,Kirsch:1994cf,gruenewald-smatasy:2005},
% \url{https://gruenew.web.cern.ch/gruenew/smatasy.html}
 originally 
developed for the analysis of LEP data in 1991/1992 and first applied in 1992 by the L3 
collaboration~\cite{sriemannL3-1233,Adriani:1993sx}. Later
applications at LEP~1, Tristan and LEP~2 are described 
in~\cite{Miyabayashi:1994ej,Yusa:1999dx,Holt:2014moa,Sachs:2003ja,ALEPH:2005ab,Schael:2013ita}.
%======================================================
%======================================================
%======================================================
% 
% See T.Riemann q
% {\bf S-matrix approach to the $Z$ line shape}
% \begin{itemize}
%  \item 
% Developed as a model-independent analysis tool of 
% \\
% \textcolor{black}{\bf $e^+e^- \to (\gamma, Z) \to f^+f^-$ around the $Z$ boson resonance  }
% %  \item 
% % In cooperation with M. Gr{\"u}newald, S. Kirsch, A. Leike, S. Riemann 
% \item 
% Aim: determinations of \textcolor{black}{$M_Z$} and \textcolor{black}{$\Gamma_Z$} 
% % in correlation with the \textcolor{black}{$\gamma Z$-interference}
% \item 
% %Refs.:
% \textcolor{black}{$\to$} $\sigma_T$: Leike/TR/Rose 1991 \cite{Leike:1991pq} 
% \\
% \textcolor{black}{$\to$} $A_{FB,LR,pol}$: TR 1992 \cite{Riemann:1992gv}, 
% \\
% \textcolor{black}{$\to$} {\tt SMATASy} code: 
% Kirsch/TR 1994 \cite{Kirsch:1994cf}
% \item 
% First application: LEP/L3 1993 \cite{Adriani:1993sx}, also: Tristan/TOPAZ,VENUS, LEP/OPAL, \ldots 
% % \item 
% % Fortran software: stand-alone \textcolor{black}{ZPOLE} (Leike/Riemann 1991, unpublished) and
% %  \textcolor{black}{SMATASY/ZFITTER} (Gr{\"u}newald/Kirsch/Riemann 1994$\to$2005) 
% % \cite{Kirsch:1994cf,Bardin:1989cw,Bardin:1990fu,Bardin:1999yd,Arbuzov:2005ma,Akhundov:2014era}
% \end{itemize}
% 
% 
% 
% The reaction
% \ba
% \textcolor{black}{e^+e^- \to (\gamma, Z) \to f^+f^- ~+~ (n~ \gamma)}
% \ea
% allows to study the $Z$ boson, its mass $M_Z$, its width $\Gamma_Z$, its couplings, and potentially deviations from the Standard Model.
% 
% \bigskip
% 
The main problem for a model-independent approach is accuracy. Even if an ansatz is improper, the fit results may nevertheless 
look 
precise:
Compare the two ansatzes for the $Z$-boson propagator, 
% {\bf Need correct ``model''}
% See experiences with \textcolor{black}{\it constant} and \textcolor{black}{\it $s$-dependent} $Z$ width:
% 
\begin{equation}
 \frac{1}{[s-M_Z^2+iM_Z~\textcolor{black}{\Gamma_Z(s)}]} ~~~~\textrm{versus}~~~~  
\frac{1}{[s-{\bar M}_Z^2+i{\bar M}_Z~\textcolor{black}{\Gamma_Z}]} .
\end{equation}
To a very good accuracy, it holds: $\Gamma_Z({s}) \approx {s}/M_Z^2 \times \Gamma_Z $. The different propagators  lead,  for one and the 
same given 
set of data, to a 
relative shift of the fitted $Z$ mass~\cite{Bardin:1988xt}:
% \bea
% {\bar M}_Z \approx M_Z - \frac{1}{2} \frac{\Gamma_Z^2}{M_Z} \approx M_Z -  34 \textrm{~MeV} .
% \eea
${\bar M}_Z \approx M_Z - {1}/{2} {\Gamma_Z^2}/{M_Z} \approx M_Z -  34 \textrm{~MeV}$ .
It is important to note that the ``wrong fit'' does not have enlarged error bars. 

\section{Total cross-sections}
The ansatz for the scattering amplitude in the complex energy plane comprises,
 in case of 
massless fermion pair production,
 { four non-interfering helicity matrix elements}:
\begin{equation}
\label{eq-smatrix}
{\bf \cal M}^i (s) = \frac {R^i_{\gamma}}{s} + \frac {R_{Z}^i}{s-s_{Z}} + F^i(s),~~~~ i=1,\cdots, 4.
\end{equation} 
The pole terms  have complex weights 
$R_{Z}$ and $R_{\gamma}$, the latter corresponding to the photon,
and the background $F(s)$ is an analytic function:
\begin{equation}
F^i(s) = \sum_{n=0}^{\infty} F_n^i (s/s_0-1)^n
\label{eq-Fs}
\end{equation}
Beware: Eqn.~(\ref{eq-smatrix}) contains the photon pole ${R^i_{\gamma}}/{s}$, where $R^i_{\gamma}$ will be assumed to have a (known) 
$s$-dependence. 
A second pole besides the $Z$ is mathematically not consistent, because a Laurent series has one single pole only.
% } {\tiny $\to$ 
% B{\"o}hm/Sato 2004 \cite{Bohm:2004zi}}
In fact, one has to understand  the term $R^i_{\gamma}(s)/s$ as part of the background term $F(s)$. Such rewritten, it reads in fact:
%{\textcolor{black}{\bf Comment on the photon term (3 Feb 2015)}}
\bea
\frac {R^i_{\gamma}(s)}{s} &=& 
% \frac{\sum_{n=0}^{\infty}R_n^i (s-s_0)^n}{s} 
% \\ 
% &=& \frac{\sum_{n=0}^{\infty}R_n^i (s-s_0)^n}{s_0-(s_0-s)} 
% \nonumber \\ 
% &=& {\sum_{n=0}^{\infty}R_n^i (s-s_0)^n}~\frac{1}{s_0} ~\frac{1}{1-\frac{s_0-s}{s_0}} 
% \nonumber \\ \nonumber
% &=& 
{\sum_{n=0}^{\infty}R_n^i (s/s_0-1)^n}~\frac{1}{s_0} ~ \left[ 1+ \frac{s_0-s}{s_0} + \left(\frac{s_0-s}{s_0}\right)^2 
\cdots\right] .
\eea
The photon pole has to be understood as part of the background terms; but once it is known as part of QED corrections, one may separate it 
from the rest of background and can take its knowledge explicitly into account.
% 
% \medskip 
% \begin{itemize}
% \item
% %\medskip 
% \textit{It is useful to sum up a selected part of the photonic background of the $Z$ resonance in order to take explicit notice of 
% physically 
% known pieces of the input expressions. }
%  \item 
% Compare: 
% \textit{It is useful to sum up a selected part of self-energy insertions in the propagators in order to derive the Breit-Wigner resonance 
% form.}
% \end{itemize}
% 
% {\textcolor{black}{\bf Ansatz for realistic applications}}

An analysis of the $Z$ line shape will use the effective Born cross section
\begin{equation}
\sigma_T(s) = \sum_{i=1}^4 \sigma^i(s) = \frac{1}{4} \sum_{i=1}^4 
            s |{\bf \cal M}^i (s)|^2.
\label{sigma}
\end{equation} 
Further, the $\sigma_T(s')$ has to be folded with a flux function in order to comprise in $\sigma_T(s)$ also QED corrections 
~\cite{Bardin:1989cw,Bardin:1990fu,Akhundov:2014era}: 
%
%-------------------------------------------from cpc
\bea
\sigma_T(s) = 
\frac{4}{3} \pi \alpha^2
\int \frac{ds'}{s} 
\left[ \frac{r^{\gamma}}{s'} +
\frac {s' {R_T} + (s' - {\bar M}_Z^2) {J_T}} {(s'-{\bar M}_Z^2)^2 + {\bar M}_Z^2 {\Gamma_Z}^2}  + \cdots
\right]
\rho^T_{ini}
\left({s'}/{s}\right).
\label{sigqed}
\eea
%-------------------------------------
Here, two real parameters besides $M_Z$ and $\Gamma_Z$ appear, the $R_T$ arising from the $Z$ pole term  and $J_T$ from the $\gamma 
Z$-interference.

The radiation from the final state may be absorbed into $\rho_{ini}^T$, and initial-final state
interferences can be taken into account by an analogue formula to
(\ref{sigqed}) with
a slightly more complicated structure: %\cite{Bardin:1990fu,Bardin:1989cw}: %B351, B229
\begin{equation}
\sigma_{T,int}(s) = \int ds' \sigma(s,s') \rho^T_{int}(s'/s).
\label{convigif}
\end{equation}
A precise description of QED, not only in the running QED coupling, but also in the flux functions $\rho_{ini}(s'/s)$ and 
$\rho_{int}(s'/s)$ is mandatory.
We mention already here that for the forward and backward cross-section parts $\sigma_F$ and $\sigma_B$ different flux functions apply, and 
so the corrections to  $\sigma_{F-B}$ are different from those to  $\sigma_{tot}$. 
Finally, it is recommended to use for the predictions of the QED corrections a sophisticated, flexible, well-tested tool like 
ZFITTER~\cite{Bardin:1989tq,Bardin:1999yd,Arbuzov:2005ma,web-sanc.zfitter:2015}. 
The recommended interface is the Fortran package SMATASY/ZFITTER~\cite{Kirsch:1994cf}. The latest version of 
SMATASY runs with ZFITTER v.6.42 
and is due to M.~Gr\"unewald~\cite{gruenewald-smatasy:2005}.
Although it is not explicitly pronounced by the authors, the copyright conditions are the same as those for ZFITTER.
Please notice  that the valid 
CPC-licence conditions~\cite{cpc-licence:7Oct2015}
%\url{http://cpc.cs.qub.ac.uk/licence/licence.html}
do not always guarantee that they are respected
 \cite{web-sanc.zfitter:2015,web-zfitter.education:2015,web-zfitter.com:2015,ombuds:2012,
zfitter-memorandum:2012,web-desy-gfitter01:2013,
web-desy-zfitter-gfitter:2014,Carminati:2014yra},
 and thus we think that it is appropriate to remind here of the importance 
%for code developers for the scientific community
 of the software authors' rights and rules of Good Scientific Practice (Appendix A).
%important for code developers for scientific community.
 
\section{Asymmetries}
\newcommand{\afb}{A_{FB}}
\newcommand{\apol}{A_{pol}}
\newcommand{\alr}{A_{LR}}
\newcommand{\rgt}{r_{\gamma}^T}
\newcommand{\rt}{R_T}
\newcommand{\rfb}{R_{FB}}
\newcommand{\rpol}{R_{pol}}
\newcommand{\ra}{R_A}
\newcommand{\rza}{r_0^A}
\newcommand{\rzt}{r_0^T}
\newcommand{\ia}{J_A}
\newcommand{\roa}{r_1^A}
\newcommand{\rot}{r_1^T}
\newcommand{\iT}{J_T}
 \newcommand{\az}{A_0}
\newcommand{\ao}{A_1}
\newcommand{\at}{A_2}
\newcommand{\st}{\sigma_T}
\newcommand{\sfb}{\sigma_{FB}}
\newcommand{\slr}{\sigma_{LR}}
\newcommand{\spol}{\sigma_{pol}}
% \beq
% \afb = \frac{\sigma_{FB}}{\sigma_T},
% \hspace{1.cm}
% \apol = \frac{\sigma_{pol}}{\sigma_T}.
% \label{e4}
% \eeq
% %------------
% {\footnotesize
% { 
% The $\afb$ and $\apol$ are helicity combinations as also $\sigma_T$ is,  i.e. have also S-matrix ansatzes.}}
% \\
% {\small The parameters in $A^{Born}(s)$ are in \textcolor{black}{[QED-]Born approximation}:}
% %------------
% \beq
% A_0 = 
% % % % \frac{ \ra + \gamma^2 \rza} { \rt + \gamma^2 (\rzt + \rgt)}
% % % % \sim
% % % % \frac{\ra}{\rt + \gamma^2 \rgt} \sim 
% \frac{{\ra}}{\rt},
% \label{e15}
% \eeq
% %------------
% {\small and}
% %------------
% {}\beq
% \ao =
% % % % \left[ \frac{J_A}{R_A} -  \frac{J_T}{R_T+\gamma^2 \rgt}
% % % %        + \frac{2 \gamma^2 \rgt}{\rt + \gamma^2 \rgt}
% % % % \right] \az % 2 \gamma^2
% % % % \sim
% \left[ \frac{{J_A}}{R_A} -  \frac{J_T}{R_T} \right] \az.
% \label{e16}
% \eeq
% %------------

% % %%%%%%%%%%%%%%%%%%%%%%%%%%%%%%%%%%%%%%%%%%%%%%%%%%%%%%%%%
% % {\textcolor{black}{\bf Scetch of derivation of the expression for $A_{FB}(s)$}}
% % 
% % % \input{smatrix-input-04.tex}
% % 
% % See \cite{RiemannT-Fccee:2015}
% % 
% %  

%%%%%%%%%%%%%%%%%%%%%%%%%%%%%%%%%%%%%%%%%%%%%%%%%%%%%%%%%%%%%%%%%%%%%%%
%[allowframebreaks=0.9]
% {\textcolor{black}{\bf QED corrections for asymmetries} }

% In the vicinity of the $Z$ boson peak,
% asymmetries behave relatively smoothly and
% may be described by a simple, universal   formula~\cite{Riemann:1992gv,Kirsch:1994cf}
Born asymmetries $\afb = {\sigma_{FB}}/{\sigma_T}$, $\apol = {\sigma_{pol}}/{\sigma_T}, \alr = {\sigma_{LR}}/{\sigma_T}$, as 
ratios of two Laurent series, are simple Taylor series. 
{QED corrections  lead to few simple correction factors; we reproduce two of them: }           % 
% \cite{sma2}:
%---------------------------------from cpc
\bea
A_{LR}^{QED}(s) &=& \phantom{c_{0,FB}(s)~} A_{0,LR}^{Born} + \textcolor{black}{c_{1,T}(s)} \,  A_{1,LR}^{Born} \left( \frac{s}{M_Z^2}-1 
\right)  + \cdots ,
\label{asyqed}
\\
A_{FB}^{QED}(s) &=& \textcolor{black}{c_{0,FB}(s)~} A_{0,FB}^{Born} + \textcolor{black}{c_{1,FB}(s)} \,  A_{1,FB}^{Born} \left( 
\frac{s}{M_Z^2}-1 \right)  + \cdots 
\eea
%-----------------------------
%\medskip
 The real constants $A_0$ and $A_1$ are of experimental relevance: 
\bea
A_{0,A} = 
% % % \frac{ \ra + \gamma^2 \rza} { \rt + \gamma^2 (\rzt + \rgt)}
% % % \sim
% % % \frac{\ra}{\rt + \gamma^2 \rgt} \sim 
\frac{{\ra}}{\rt},
~~~
% \label{e15}
% \eea
% {}\bea
A_{1,A} =
% % % \left[ \frac{J_A}{R_A} -  \frac{J_T}{R_T+\gamma^2 \rgt}
% % %        + \frac{2 \gamma^2 \rgt}{\rt + \gamma^2 \rgt}
% % % \right] \az % 2 \gamma^2
% % % \sim
\left[ \frac{{J_A}}{R_A} -  \frac{J_T}{R_T} \right] \az,~~~ A=FB,pol,LR.
\label{e16}
\eea
The QED corrections are contained in smooth set-up dependent, {\it model-independent} factors \textcolor{black}{$C(s)$}:
%--------------------------------------------from desy 94-125, july 1994, cpc
% Licensee: Tord  Riemann
% License Date: Jan 27, 2015
% License Number: 3557090997554
\bea
\textcolor{black}{c_{0,FB}(s)}= \frac{C_{FB}^R}{C_T^R}, ~~~
\textcolor{black}{c_{0,T}(s)}=1 
, ~~~  %\\
\textcolor{black}{c_{1,A}(s)}=c_{0,A}~\frac{C_T^J}{C_T^R},~~~ a=T,FB.
\eea
% \vspace*{-4mm}
As examples, we reproduce $C_{T}^R(s)$ and $C_{FB}^R(s)$:
\bea
C_{\textcolor{black}{A}}^R(s)
&=&
\int dk ~{\textcolor{black}{\rho^{A}_{ini}(s'/s)}}\frac{s'R_A}{sR_A} \frac{(s-{\bar M}_Z^2)^2 + {\bar M}_Z^2 \Gamma_Z^2}{(s'-{\bar 
M}_Z^2)^2 + {\bar M}_Z^2 \Gamma_Z^2},~~~ a=T,FB.
\eea

% 
% 
% 
% 
% 
% %%%%%%%%%%%%%%%%%%%%%%%%%%%%%%%%%%%%%%%%%%%%%%%%%%%%%%%%%
% %[allowframebreaks=0.9]
% {\textcolor{black}{\bf Scetch of derivation of the expression for $A_{FB}(s)$ with QED corr's}
% }
% 
% % \input{smatrix-input-05.tex} 
% 
% See \cite{RiemannT-calc:2015}
% 
% 
% 
% 
% 
% 
% 
% 
% %[allowframebreaks=0.9]
% {\textcolor{black}{\bf QED corrections for asymmetries} }

% 
% \begin{figure}[thbp]
% %---
% \begin{center}
% \vspace*{-10mm}
% \includegraphics[scale=0.35]{asy3.pdf}
% %\mbox{\epsfxsize=\swid\epsffile {asy.eps}}
% \end{center}
% \vspace*{-12mm}
% \caption[foo]{
% The forward-backward asymmetry for the process $e^+ e^- \rightarrow \mu^+
% \mu^-$ near the $Z$ boson peak. From Kirsch/Riemann 1994 \cite{Kirsch:1994cf}, license Number: 3557090997554.
% \label{asy}
% }
% \end{figure}
% %------------------------------------------------------------------

\section{Applications}
% \subsection{} %trie this command activates in the headline the dots for pages
% %[allowframebreaks=0.9]
% {\textcolor{black}{\bf Applications} }
The described model-independent approach allows experimental fits to the mass and width of the $Z$ boson which are, in accuracy, 
competitive to the Standard Model approach. 
The minimal number of data points (in $s$) will be five, as may be seen from (\ref{sigqed}). 
There is a correlation of $Z$~peak position $s_{peak}$, the $Z$ mass value, and the $\gamma Z$ 
interference term 
$J_T$:
{}
%------------
\bea
\Delta \sqrt{s_{peak}} = \Delta {\bar M}_Z + \frac{1}{4} \frac{\Gamma_Z^2}{{\bar M}_Z} \, \,
\Delta    \left( \frac{\iT}{\rt} \right).
\label{e17}
\eea
%------------
Any misidentification of $J_T$ leads to a correlated systematic shift of ${\bar M}_Z$. In the Standard model, the $J_T$ is  a derived 
quantity and thus fixed, while 
here it is a floating quantity - if we do not decide to fix it ``by hand'' or by other data.  
As a consequence, the error bars in the strict S-matrix approach for ${\bar M}_Z$ will be  systematically a bit bigger than in the SM 
approach.
For further discussions, we refer to the literature quoted and to \cite{Leike:1992uf,sriemannL3-1233,Grunewald:1993zw,RiemannT-calc:2015}.
Whether the S-matrix approach can be useful at the FCC-ee deserves a detailed investigation on the approximations made in the 
SMATASY/ZFITTER toolkit which have not been discussed here.

\section*{Acknowledgements}
I would like to thank M. Gr{\"u}newald and S. Riemann for discussions.
With pleasure I remember the collaboration with M. Gr{\"u}newald, Arnd Leike, Stefan Kirsch, Sabine Riemann, Joachim Rose in different 
stages of the project. I thank Alain Blondel and Fulvio Piccinini for the  invitation to review the project for the FCC-ee. 

This work is supported by the Polish National Science Centre
(NCN) under the Grant Agreement No. DEC-2013/11/B/ST2/04023.
%%%%%%%%%%%%%%%%%%%%%%%%%%%%%%%%%%%%%%%%%%%%%%%%%%%%%%
%\clearpage
\appendix
\section{Good Scientific Practice and software}
As quoted above, it happens that the rules of Good Scientific Practice are violated when software is concerned.
The reasons root deeply in history. For decades, software was not considered as a genuine result of scientific work, but mere 
as an auxiliary work with no value by itself. This changed with its rising complexity. But until today, in almost all of the official 
documents on Good Scientific Practice there is mentioning of texts, figures, ideas, data, but no mentioning of software. And a need of its 
protection against misuse - by lack of attribution or by improper use  -  is questioned, especially by non-experts .

International academic fundamental research relies on universal principles and ethical rules, and on national legal regulations.
%; the latter are not international. 
We are seekers of the truth.
One of our most important principles is honesty. 
Society has to trust us in our doings because an independent control is, due to the complexity of our work, impossible. 
We researchers carry the responsibility to prevent and, in case it happens, sanction fraud in science.
Fraud destroys the balance of competition and cooperation. It also destroys the trust, and it finally destroys the contract of society and 
science.

Academic research, and we are part of that, is distinguished from other research by the need of correct attribution:
\begin{itemize}
\item {\em Attribution} of a scientific achievement to those who made it.
%\item 
\\ Often by a proper citation.
%\item 
%\\ 
Other proper attributions are possible.
\end{itemize}
There is no need to explain why proper attribution is substantial. Violations are called plagiarism.
A proper attribution informs on:
\begin{itemize}
\item {\em What was done?} ~~
% \item \\
{\em Who did it?} ~~
%\item \\
{\em How important is it?}
\end{itemize}
In case of ethical or legal problems, seek cooperations and not confrontations. Questions have to be answered, in this order:
\begin{itemize}
\item {\em Facts:} What are the initial facts? Investigate carefully.
%\item 
\\
{\em Rules:} Are there violated rules? Seek a healing by negotiations. 
%\item 
\\
{\em Sanctions:} In case. Are there sanctions foreseen? By whom?
\end{itemize}
A round table discussion on ``Open-source, knowledge sharing
and scientific collaboration'' at ACAT2013 is summarized in~\cite{Carminati:2014yra}. % also: yra
An important point is the authors' right to set ``conditions of use'', often formalized in licence agreements.  In scientific practice, 
they replace law - if they are respected. 
We use here the CPC-licence ~\cite{cpc-licence:7Oct2015}, which is 
 unfortunately questioned by a 
variety of colleagues and institutions. If this will become a common practise
 no scientist will be willing to develop a software to be shared
  by the community.  % and destabilizes the science system. 
Other well-known licence model families are GPL licences (Gnu public licences, not ideal for academic attributions, although often 
recommended for software) and CC licences (Creative Commons licences). 
To respect licences as valid ``conditions of use'' is essential because there is no international copyright law.  With no 
doubt -- it would be extremely difficult to authors to defend rights at a court. The big international scientific centers and all the 
national science organizations are asked not only to set the rules, but also to defend them.

\bigskip

%%% note added 2016-10-14 =======
{\bf Note added on 14 October 2016 for submission to the hep-ph arxive}

\medskip

\begin{quote}
All the people we used to know  \\
They're an illusion to me now  \\
Some are mathematicians  \\
Some are carpenters' wives  \\
Don't know how it all got started  \\
I don't know what they're doin' with their lives  \\
But me, I'm still on the road  \\
Headin' for another joint  \\
We always did feel the same  \\
We just saw it from a different point  \\
Of view  \\
Tangled up in blue  \\
\end{quote}
Text \copyright ~Bob Dylan Music Co. (1974) \href{http://bobdylan.com/songs/tangled-blue}{http://bobdylan.com/songs/tangled-blue}

\bigskip

The present text has been published in the proceedings of the Ustron meeting "Matter to the deepest" in 
Acta Physica Polonica vol. 46 (2015) No 11, p. 2235, with the affiliation K{\"o}nigs Wusterhausen, Germany. This is one of the places 
where I live and work. 

\medskip

In 2015, I gave four quite similar presentations of the topic at meetings at Pisa, Dubna, Ustron, Karlsruhe.
For the presentation at Karlsruhe [3] I asked, on 2 October 2015, the scientific representative of DESY Dr. H. Dosch for the approval of 
the slides.
I got the written approval to give the talk (as it was submitted) by email on 7 October 2015:
"\ldots im Auftrag von Herrn Dosch und in Abstimmung mit Herrn Mnich richte ich
Ihnen hiermit die Genehmigung Ihrer Slides aus."
Unfortunately, the approval was given the day after the talk, and thus there are two versions of it at the conference webpage, with 
different affiliations,
\href{https://indico.cern.ch/event/369827/contributions/875702/attachments/1166637/1682217/2-riemann_tord.pdf}{
https://indico.cern.ch/event/369827/contributions/875702/attachments/\linebreak[4]1166637/1682217/2-riemann\_tord.pdf}
and 
\href{https://indico.cern.ch/event/369827/contributions/875702/attachments/1166637/1683306/2-riemann_tord_backup.pdf}{
https://indico.cern.ch/event/\linebreak[4]369827/contributions/875702/attachments/1166637/1683306/2-riemann\linebreak[4]\_tord\_backup.pdf}.

\medskip

On 12 October 2015, I asked the scientific representative of DESY for the permission to publish the proceedings version of the slides as a 
DESY Red Report (filename: 2015-ustron-proc-riemann-sMatrix-rR.pdf, 13 Oct 2015).
The text was not accepted for publication (Dr. J. Mnich, 18.11.2015), and on 6 December 2015 I appealed to the ad-hoc publication commission 
of DESY (Dr. M. K{\"o}hler, Dr. C. Stegmann, Dr. M. Diehl, established on 17 December 2015).
The commission's report is dated four months later, on 21 April 2016.

The commission decided that the present text would be appropriate for publication if certain changes would be undertaken.
I refused, and the text did not get the approval as a DESY Red Report. 
%Below I reproduce the relevant contents of the Report of the DESY publication commission.

\medskip

The contents of my talk on "S-matrix approach to the Z resonance" summarizes the publications on the S-Matrix approach to the Z resonance 
from 1991 till now. The motivation is to have a mini-review for later use with respect to physics studies for the ILC.
With the ILC, one aims at per mille accuracy of many observables, and many of the analysis tools have to be re-analyzed therefore. 
Including the S-matrix analysis tools.

In the meantime, we finished (in August 2016) a 2-loop project on the weak mixing angle in the standard model related to the Zbb vertex; 
see:
I. Dubovyk, A. Freitas, J. Gluza, T. Riemann, J. Usovitsch, PLB 762 (2016) 184, 
\href{http://dx.doi.org/10.1016/j.physletb.2016.09.012}{http://dx.doi.org/10.1016/j.physletb.2016.09.012}, preprint KW 16-002, 
\href{http://arxiv.org/abs/1607.08375}{http://arxiv.org/abs/1607.08375}.
\\
I could not understand one point -- how to get the pseudo-observable $A_{FB}$, and thus $\sin^2\theta_W^{b}$, from realistic cross 
sections, when observing the correct ansatz for the weak amplitude in the standard model as demanded by analyticity, causality, unitarity 
and gauge invariance.     
And then I understood: The answer is the S-matrix approach to realistic observables.
\\
Not interpreted as an alternative to perturbation theory, as I did so far, but in conformity with perturbation theory.

\bigskip

Why did I not make the text changes in the manuscript in fulfillment of the proposals of the DESY commission?

\medskip 

1. References [21,23-29] were demanded to get shortened to [24,30,31]. (The references in the commission's letter refer 
to a text slightly different from the APP text.)
The change might have been tolerable.

\medskip 

2. Webpage [21] \href{http://sanc.jinr.ru/users/zfitter}{http://sanc.jinr.ru/users/zfitter} was demanded to be replaced by the archived 
webpage \href{http://dx.doi.org/10.3204/zfitter.education}{http://dx.doi.org/10.3204/zfitter.\linebreak[4]education} held at the DESY 
domain.
\\
The change is not acceptable.
\\
The ZFITTER webpage at JINR is the project page of the ZFITTER project, and it has to be dynamic as the project is dynamic, and it has to 
be under the authorship of the project authors. 

\medskip 

3. Ref. [29]: Carminati, Perret-Gallix, Riemann, "Summary of the ACAT 2013 round table discussion: Open-source,                         
knowledge sharing and scientific collaboration", J. Phys. Conf. Ser. 523, 012066, 
\href{http://dx.doi.org/10.1088/1742-6596/523/1/012066}{http://dx.doi.org/10.1088/1742-6596/523/1/012066}, 
\href{http://arxiv.org/abs/1407.0540}{http://arxiv.org/abs/\linebreak[4]1407.0540}, Red Report DESY 14-079. This reference  is demanded to 
be omitted.
\\
The reason is not made clear.

\medskip 

I did not fulfill demands 1. to 3. of the DESY publication commission, and the publication was not approved by the scientific 
representative of DESY. My written appeal to the scientific representative of DESY to ignore the recommendations of the DESY publication 
commission was not answered.

As a result of the developments described to my best knowledge, the publication was not approved with a DESY affiliation.

%%% note added 2016-10-14 end =======

% % % \providecommand{\href}[2]{#2}
% % % %\bibliographystyle{JHEP-2} % also: {JHEP}
% % % %\bibliographystyle{elsarticle-num} % for 2015 talk, mit hep-ph + Titel + doi
% % % \bibliographystyle{utphys_spires} % for 2013 proceedings, mit hep-ph, ohne Titel und ohne doi
% % % \bibliography{2loops}

\providecommand{\href}[2]{#2}

\end{document}